\documentstyle[12pt,a4,psfig,axodraw]{article}
\def\bite{\begin{itemize}}
\def\eite{\end{itemize}}

\def\PR{\frac{1 + \gamma_5}{2}}
\def\PL{\frac{1-\gamma_5}{2}}
\def\hc{{\rm h.c.}}
\def\mstone{\wt{M}_{t1}}
\def\msttwo{\wt{M}_{t2}}
\def\mg{m_{3/2}}
\def\stopone{\wt{t}_1^{}}
\def\stoptwo{\wt{t}_2^{}}
\def\stopl{\wt{t}_L^{}}
\def\stopr{\wt{t}_R^{}}
\def\MQ{\wt{m}_Q}
\def\MU{\wt{m}_U}

\def\sn2w{\sin^2\theta_W^{}}
\def\chargino{\omega}

\def\diag{{\rm diag}}
\def\half{\frac{1}{2}}
\def\vone{v_1^{}}
\def\vtwo{v_2^{}}
\def\rttwo{\sqrt{2}}

\def\exp{{\rm exp}}
\def\wt{\widetilde}
\def\mgaugino{\wt{m}}
\def\mhiggsino{m_H^{}}
\def\mhiggsinos{m_H^*}

\def\acp{{\cal A}_{CP}}
\def\bsgamma{b \to s \gamma}
\def\Bsgamma{B \to X_s \gamma}
\def\to{\rightarrow}

\def\vsk#1{\noalign{\vskip#1 cm}}
\def\vsp#1{\vspace{#1 cm}}

\def\ov{\overline}

\def\gsim{~{\rlap{\lower 3.5pt\hbox{$\mathchar\sim$}}\raise 1pt\hbox{$>$}}\,}
\def\lsim{~{\rlap{\lower 3.5pt\hbox{$\mathchar\sim$}}\raise 1pt\hbox{$<$}}\,}

\def\etal{{\it et al.}}
\newcommand{\beq}{\begin{equation}}
\newcommand{\eeq}{\end{equation}}
\newcommand{\bea}{\begin{eqnarray}}
\newcommand{\eea}{\end{eqnarray}}
\newcommand{\bsub}{\begin{subequations}}
\newcommand{\esub}{\end{subequations}}
\renewcommand{\theequation}{\thesection.\arabic{equation}}
\newcommand{\clean}{\setcounter{equation}{0}}

\def\PRD#1#2#3{Phys. Rev. {\bf D#1} (19#2) #3}
\def\NPB#1#2#3{Nucl. Phys. {\bf B#1} (19#2) #3}
\def\PTP#1#2#3{Prog. Theor. Phys. {\bf #1} (19#2) #3}

\def\EPJC#1#2#3{Eur. Phys. J. {\bf C#1} (19#2) #3}
\def\PLB#1#2#3{Phys. Lett. {\bf B#1} (19#2) #3}
\def\PRL#1#2#3{Phys. Rev. Lett. {\bf #1} (19#2) #3}
\def\PRep#1#2#3{Phys. Rep. {\bf #1} (19#2) #3}
\makeatletter
\newtoks\@stequation

\def\subequations{\refstepcounter{equation}%
  \edef\@savedequation{\the\c@equation}%
  \@stequation=\expandafter{\theequation}
  \edef\@savedtheequation{\the\@stequation}
  \edef\oldtheequation{\theequation}%
  \setcounter{equation}{0}%
  \def\theequation{\oldtheequation\alph{equation}}}

\def\endsubequations{%
  \ifnum\c@equation < 2 \@warning{Only \the\c@equation\space subequation
    used in equation \@savedequation}\fi
  \setcounter{equation}{\@savedequation}%
  \@stequation=\expandafter{\@savedtheequation}%
  \edef\theequation{\the\@stequation}%
  \global\@ignoretrue}

\def\eqnarray{\stepcounter{equation}\let\@currentlabel\theequation
\global\@eqnswtrue\m@th
\global\@eqcnt\z@\tabskip\@centering\let\\\@eqncr
$$\halign to\displaywidth\bgroup\@eqnsel\hskip\@centering
     $\displaystyle\tabskip\z@{##}$&\global\@eqcnt\@ne
      \hfil$\;{##}\;$\hfil
     &\global\@eqcnt\tw@ $\displaystyle\tabskip\z@{##}$\hfil
   \tabskip\@centering&\llap{##}\tabskip\z@\cr}

\makeatother

\begin{document}
\begin{titlepage}
\vspace*{-15mm}


\baselineskip 10pt
\begin{flushright}
\begin{tabular}{l}
{\bf OCHA-PP-133} \\
{\bf KEK-TH-614}\\
{\bf ICRR-Report-451-99-9}\\
\end{tabular}
\end{flushright}
\baselineskip 24pt 
\vglue 10mm 
\begin{center}
{\Large\bf
CP asymmetry for radiative $B$-meson decay in the supersymmetric 
standard model 
}
\vspace{5mm}

\baselineskip 18pt 
\def\thefootnote{\fnsymbol{footnote}}
\setcounter{footnote}{0}
{\bf
Mayumi Aoki$^{1}$\footnote{Research Fellow of the Japan Society 
for the Promotion of Science.}, 
Gi-Chol Cho$^{2*}$, and Noriyuki Oshimo$^3$
}
\vspace{5mm}

$^1${\it Graduate School of Humanities and Sciences, 
Ochanomizu University \\
Bunkyo-ku, Tokyo 112-8610, Japan}\\
$^2${\it Theory Group, KEK, Tsukuba, Ibaraki 305-0801, Japan}\\
$^3${\it Institute for Cosmic Ray Research, University of Tokyo \\
Tanashi, Tokyo 188-8502, Japan}\\
\vspace{12mm}
\end{center}

\begin{center}
{\bf Abstract}\\[7mm]
\begin{minipage}{13cm}
\baselineskip 18pt
\noindent
We study a rate asymmetry in the inclusive decay $\Bsgamma$,  
assuming the supersymmetric standard model based on $N=1$ supergravity.  
A complex coefficient $A$ for scalar-trilinear couplings is 
the source of CP violation, 
which is contained in the mass-squared matrices for squarks.  
The model parameters are nontrivially constrained by 
experimental results for the branching ratio 
of the radiative decay and for 
the electric dipole moments of the neutron and the electron.  
The decay rate asymmetry is predicted to be much larger than 
that by the standard model in a wide region of the parameter space 
compatible with the experiments.  
Its magnitude can be maximally around 0.1, which will be 
well accessible at B factories. 

\end{minipage}
\end{center}
\vsp{0.15}
 
{\small 
 \begin{flushleft}
 \end{flushleft}
 }
 
\end{titlepage}

\newpage
\baselineskip 18pt 
\def\thefootnote{\arabic{footnote}}
\setcounter{footnote}{0}
\def\r2{\sqrt 2}
\def\w{\omega}
\def\HC{H^\pm}
\def\M{\widetilde M}

%
%
%
\section{Introduction}
\clean

It has been considered that CP violation in the $B$-meson system may 
give an important clue to physics beyond the standard 
model (SM) \cite{nir_quinn}.  
At B factories such CP asymmetries will be measured in various reactions.  
One possible candidate is the process $B_d \to \psi K_S$, 
which will be observed with high statistics.  
In this process the $b$-quark decay occurs 
dominantly at the tree level by the SM interaction, 
and new physics could give contributions through $B^0$-$\ov{B^0}$ mixing.  
Another possibility is in the inclusive decay $\Bsgamma$, 
where the $b$ quark decays at the one-loop level.  
Although its statistics are not high, new physics could affect  
the $b$-quark decay.  

In this paper, we study a decay rate asymmetry in $\Bsgamma$  
within the framework of the supersymmetric standard model (SSM) 
based on $N=1$ supergravity and grand unified theories (GUTs)~\cite{SUGRA}. 
This CP asymmetry is defined as the difference of the decay rates 
between $\ov{B}\to X_s\gamma$ and its CP conjugate process: 
\bea
\acp &=& \frac{\Gamma({\ov B} \to X_{s} \gamma)- \Gamma(B \to X_{\ov s} \gamma)}
{\Gamma({\ov B} \to X_{s} \gamma)+ \Gamma(B \to X_{\ov s} \gamma)}. 
\label{eq:asymmetry}
\eea
The SM prediction is less than 0.01~\cite{soares}. 
The SSM contains new interactions for the $b$ quark which 
could sizably induce processes of flavor-changing 
neutral current (FCNC) \cite{oshimo,branco_cho_kizukuri_oshimo} 
and of CP violation \cite{aoki_oshimo}.    
We can expect an enhancement of the CP asymmetry for $\Bsgamma$, 
which indeed has been shown recently~\cite{aoki_cho_oshimo}.  
On the other hand, the new interactions for the quarks 
affect the branching ratio 
of the radiative decay and the electric dipole moments (EDMs) of the neutron 
and the electron, both of which have been measured in experiments 
rather precisely.    
Taking into account these experimental constraints, 
we discuss the asymmetry and analyze its dependencies on 
various SSM parameters.   
It will be shown that the asymmetry has a magnitude 
larger than the SM prediction 
in a wide region of the parameter space allowed by the experiments.  
Supersymmetry may be revealed through the decay rate asymmetry 
for $\Bsgamma$ at B factories.  

The decay $\Bsgamma$ is well described by the quark-level processes 
$\bsgamma$ and $\bsgamma g$, owing to a large mass of the $b$ quark.  
In the SSM, there are several new sources of  
FCNC which give rise to these decays at the one-loop level. 
In particular, the interactions of down-type quarks with 
charginos and up-type squarks and with charged Higgs bosons 
and up-type quarks play dominant roles~\cite{oshimo}.  
Furthermore, the former interactions violate CP 
invariance \cite{aoki_oshimo}, which is attributed 
to the chargino mass matrix or the $t$-squark mass-squared matrix.  
We study the effects of these interactions on the CP asymmetry.  
Although the interactions of down-type quarks with 
gluinos and down-type squarks or with neutralinos and down-type squarks 
could also become sources of FCNC and CP violation, 
their effects are suppressed by small off-diagonal 
elements of the $b$-squark mass-squared matrix.  

Experiments have put stringent bounds on the branching ratio 
of $\Bsgamma$ and the EDMs of the neutron and the electron.   
For the branching ratio, two measurements have been reported 
by CLEO \cite{cleo} and ALEPH \cite{aleph} to give respectively  
\bsub
\bea
{\rm Br}(\Bsgamma) &=& (3.15 \pm 0.35 \pm 0.32 \pm 0.26)
	\times 10^{-4}, 
\\
                 &=& (3.11 \pm 0.80 \pm 0.72) \times 10^{-4}.   
\eea
\esub
Both results are consistent with the prediction of the SM including  
next-to-leading order (NLO) corrections for QCD, 
${\rm Br}(\Bsgamma) = (3.29\pm 0.33)\times 10^{-4}$ \cite{kagan_neubert2}.  
The experimental upper bounds on the neutron and the electron EDMs 
have been obtained as $|d_n|\lsim 10^{-25}e$cm \cite{neutron-edm} 
and $|d_e|\lsim 10^{-26}e$cm \cite{electron-edm}.  
The SM predictions are given by 
$|d_n|=10^{-33}-10^{-31}e$cm \cite{neutron-edm-rev} 
and $|d_e|\lsim 10^{-37}e$cm \cite{electron-edm-rev}.  
These experimental results lead to useful constraints on SSM parameters.  

This paper is organized as follows.   
In Section~2, we summarize sources of CP violation and FCNC 
in the SSM and present the interaction Lagrangians which are relevant 
to the decay $\Bsgamma$. 
In Section~3, Wilson coefficients for the radiative decay 
are obtained and formulae for the CP asymmetry are given  
explicitly in a self-contained form.  
In Section 4, numerical results for the CP asymmetry are presented  
together with the branching ratio.  
Discussions and conclusions are given in Section~5. 

%
%
%
\section{Sources of CP violation and FCNC}
\clean

In the SSM coupled to $N=1$ supergravity, 
there are many complex parameters,  
as well as Yukawa coupling constants: 
a mass parameter $\mhiggsino$ in the bilinear term of the Higgs 
superfields, SU(3), SU(2), 
and U(1) gaugino masses $\mgaugino_3$, $\mgaugino_2$, 
and $\mgaugino_1$, and dimensionless coefficients $A_f$ and $B$ 
for scalar-trilinear and -bilinear couplings, 
$f$ representing a flavor for quarks and leptons.  
Assuming that the gaugino masses  
have a common value at around the GUT scale,  we take 
\beq
\frac{3\mgaugino_1}{5\alpha_1} = \frac{\mgaugino_2}{\alpha_2} 
= \frac{\mgaugino_3}{\alpha_3}   
\label{eq:gauginomass}
\eeq
at the electroweak scale.  
We also assume that the differences of $A_f$ among flavors are small, 
putting the same value on them $A_f=A$.  
Then only two new complex phases become physical.  
By appropriate field redefinitions,  
we can take without loss of generality $\mhiggsino$ and $A$ as complex,  
\beq
\mhiggsino = |\mhiggsino| \exp(i\theta), \quad 
A = |A| \exp(i\alpha), 
\label{eq:cp_sources}
\eeq
and the gaugino masses and $B\mhiggsino$ as real.  
The vacuum expectation values $\vone$ and $\vtwo$ of the Higgs 
bosons with hypercharges $Y=-1/2$ and $+1/2$, respectively, 
become real in this convention.  

The new source of FCNC relevant to $\Bsgamma$ is the 
mass-squared matrix $M_U^2$ for up-type squarks, 
which contains the physical complex phases $\alpha$ and 
$\theta$ intrinsic in the SSM 
and thus also becomes an origin of CP violation. 
Since the squarks are mixed not only among different generations 
but also between left- and right-handed components, 
$M_U^2$ is expressed by a $(6\times6)$ matrix.  
We assume that  
the generation mixings in $M_U^2$ can be removed approximately by 
the same unitary matrices which diagonalize the mass 
matrix of up-type quarks.   
Then, the generation mismatches between up-type squarks 
and down-type quarks in the interactions with charginos 
are described by the Cabibbo-Kobayashi-Maskawa (CKM) matrix 
for the quarks.  

The magnitude of the left-right mixing for the squarks $\wt{q}_k$ 
is proportional to the Yukawa coupling constant for 
the corresponding quark $q$. 
Consequently, in the first two generations, the left-right mixings
are safely neglected.  
The left-handed squarks $\wt{u}_L, \wt{c}_L$ and the right-handed squarks 
$\wt{u}_R, \wt{c}_R$ are themselves in  mass eigenstates, 
whose masses are given by    
\bsub
\bea
\wt{M}_{uL}^2 &=& \wt{M}_{cL}^2 = 
	\MQ^2 + \cos2\beta\biggl( \half - \frac{2}{3}\sn2w \biggr) M_Z^2, 
\\
\wt{M}_{uR}^2 &=& \wt{M}_{cR}^2 = 
	\MU^2 + \frac{2}{3} \cos2\beta \sn2w M_Z^2. 
\eea
\esub
Here $\MQ$ and $\MU$ denote soft supersymmetry-breaking masses 
for the squarks of SU(2) doublets and singlets, respectively, 
and have roughly the same value.  
The angle $\beta$ is defined by $\tan\beta = \vtwo/\vone$. 
The $u$- and $c$-quark masses have been neglected. 

For the squarks of the third generation $\stopl$ and $\stopr$, 
the left-right mixing is no longer negligible,  
because of a large Yukawa coupling constant for the $t$ quark .   
The mass-squared matrix $M_t^2$ for the $t$ squarks 
is given by 
\bea
M_t^2 &=& \left(
	\begin{array}{cc}
	\wt{M}_{uL}^2 + (1-c)m_t^2 & m_t(A^* \mg + \cot\beta \mhiggsino) 
\\
\vsk{0.2}
	m_t(A \mg + \cot\beta \mhiggsinos ) & 
			\wt{M}_{uR}^2 + (1-2c) m_t^2 
	\end{array}
	\right), 
\eea
where $\mg$ represents the gravitino mass, satisfying 
$\MQ\approx\MU\approx|A|\mg$.   
A dimensionless constant $c$ is introduced to parametrize  
radiative corrections to the squark masses through Yukawa interactions,   
with $c=0.1-1$.  
The mass eigenstates of the $t$ squarks $\stopone$ and $\stoptwo$ 
are obtained by diagonalizing $M_t^2$ as 
\bea
S_t^\dagger M_t^2 S_t &=& \diag( \mstone^2, \msttwo^2 )
~~~~~~~~~( \mstone^2 < \msttwo^2 ), 
\eea
where $S_t$ is a unitary matrix. 

The CP-violating phase $\theta$ also  
appears in the mass matrix for charginos $\chargino_i$, 
which are charged mass eigenstates of SU(2) gauginos and Higgsinos.   
The mass matrix is given by 
\bea
M^- &=& \left( 
	\begin{array}{cc}
	\mgaugino_2 & -\rttwo\cos\beta M_W \\
	-\rttwo\sin\beta M_W & \mhiggsino
	\end{array}
	\right). 
\eea
The mass eigenstates are obtained by diagonalizing $M^-$ as 
\bea
C_R^\dagger M^- C_L &=& \diag( m_{\w 1}, m_{\w 2})
~~~~~~~~( m_{\w 1} <  m_{\w 2}), 
\eea
where $C_R$ and $C_L$ are unitary matrices. 

The interaction Lagrangians which could give large new  
contributions to the decays $\bsgamma$ and $\bsgamma g$ are given 
as follows \cite{oshimo,branco_cho_kizukuri_oshimo}: 
\begin{flushleft}
{\it The chargino - quark - squark interactions}
\end{flushleft}
\bea
{\cal L} &=& 
	i\frac{g}{\rttwo} \sum_{i=1}^2 
	\Biggl\{ 
	\biggl( 
		\wt{u}_L^\dagger, \wt{c}_L^\dagger, 
		\sum_{k=1}^2 S_{t1k}^* \wt{t}_k^\dagger
	\biggr)  
	V \ov{\chargino}_i 
	\biggl (
	\rttwo C_{R1i}^* \PL 
		+ \frac{\ov{m}_D}{M_W} \frac{C_{L2i}^*}{\cos\beta} \PR 
	\biggr)
	\left( \begin{array}{c}
	d \\ s \\ b \end{array} \right)
\nonumber \\
&-& 
	\biggl( 
	\wt{u}_R^\dagger, \wt{c}_R^\dagger, 
	\sum_{k=1}^2 S_{t2k}^* \wt{t}_k^\dagger \biggr) 
	\frac{C_{R2i}^*}{\sin\beta} 
	\frac{\ov{m}_U}{M_W} V \ov{\chargino}_i \PL 
	\left( \begin{array}{c}
	d \\ s \\ b \end{array} \right)
	\Biggr \} + \hc,  
\label{chargino lagrangian}
\eea
\begin{flushleft}
{\it The charged Higgs boson - quark - quark interactions}
\end{flushleft}
\bea
{\cal L} &=& \frac{g}{\rttwo} 
H^+ \left( \begin{array}{ccc} 
	\ov{u},&  \ov{c},&  \ov{t}
	\end{array} \right )
\biggl(
	\cot\beta \frac{\ov{m}_U}{M_W} V \PL 
	+ \tan\beta V\frac{\ov{m}_D}{M_W}  \PR
\biggr) 
	\left( \begin{array}{c}
	d \\ s \\ b \end{array} \right) 
\nonumber \\
&& + \hc.  
\label{higgs lagrangian}
\eea
Here $V$ stands for the CKM matrix and  
$\ov{m}_U$ and $\ov{m}_D$ represent the diagonalized quark mass matrices
\bea
\ov{m}_U &=& \diag(0, 0, m_t^{}), 
~~~~
\ov{m}_D = \diag(0, 0, m_b^{}), 
\eea
with the quark masses of the first two generations being neglected.  

Rigorous constraints on the new CP-violating phases $\theta$ and $\alpha$  
come from the EDMs of the neutron and the electron. 
The EDMs receive contributions from one-loop diagrams mediated by    
squarks or sleptons with gluinos, charginos, or neutralinos.   
If CP violation by the phase $\theta$ is not suppressed, 
the experimental bounds on the EDMs constrain  
squark and slepton masses to be larger than 1 TeV~\cite{kizukuri_oshimo}.  
On the other hand, the phase $\alpha$ is allowed to maximally cause  
CP violation even if squarks and 
sleptons are not heavy \cite{aoki_oshimo,aoki_sugamoto_oshimo}.  
Only the gluinos are required to be sufficiently heavy, 
leading to $\mgaugino_2\gsim 500$ GeV from Eq. (\ref{eq:gauginomass}).  
The decay rate asymmetry in Eq. (\ref{eq:asymmetry}) is expected 
to have a large magnitude only if at least one CP-violating phase 
is not suppressed and the charginos and the squarks are not heavy.  
We therefore assume that $\theta$ is about 0 or $\pi$, while 
no constraint is imposed on $\alpha$.    

%
%
%
\section{Decay rate asymmetry and branching ratio}
\clean

The radiative decay $\Bsgamma$ can be approximated  
by the free quark processes $\bsgamma$ and $\bsgamma g$. 
These elementary processes for the $\Delta B = 1$ transition 
are studied by using the effective Hamiltonian with five quarks 
in which heavier degrees of freedom are integrated out:  
\bea
{\cal H}_{eff} &=& -\frac{4G_F}{\rttwo} V_{ts}^* V_{tb} 
	\sum_{j=1}^8 C_j(\mu) O_j(\mu). 
\eea
Here $O_j(\mu)$ and $C_j(\mu)$ represent an operator for the 
$\Delta B = 1$ transition and its Wilson coefficient, respectively,  
evaluated at an energy scale $\mu$. 
The relevant operators for $\Bsgamma$ are given by 
\bsub
\bea
O_2 &=& \ov{s_L} \gamma^\mu c_L \ov{c_L} \gamma_\mu b_L, \\
O_7 &=& \frac{e}{16\pi^2} m_b^{} \ov{s_L} \sigma^{\mu\nu} 
	b_R F_{\mu\nu}, \\
O_8 &=&  \frac{g_s}{16\pi^2} m_b^{} \ov{s_L} \sigma^{\mu\nu} 
	T^{\alpha} b_R G^{\alpha}_{\mu\nu}, 
\eea
\esub
where $F_{\mu\nu}$ and $G^{\alpha}_{\mu\nu}$ respectively 
represent the electromagnetic 
and the strong field strength tensors, $T^\alpha$ being a generator 
for SU(3).   
The CP asymmetry and branching ratio for $\Bsgamma$ are expressed 
in terms of the Wilson coefficients at $\mu=m_b$.   

\begin{figure}
\begin{center}
\begin{picture}(400,150)(0,0)
\Line(140,20)(260,20)
\PhotonArc(200,20)(30,0,180){5}{8}
\Text(130,20)[]{\large $b$}
\Text(200,60)[]{\large $W$}
\Text(200,8)[]{\large $u,c,t$}
\Text(270,20)[]{\large $s$}
\Line(40,100)(70,100)
\CArc(100,100)(30,0,180)
\DashLine(70,100)(130,100){5}
\Line(130,100)(160,100)
\Text(30,100)[]{\large $b$}
\Text(100,140)[]{\large $\omega_i$}
\Text(100,88)[]{\large $\tilde u_k,\tilde c_k,\tilde t_k$}
\Text(170,100)[]{\large $s$}
\Line(240,100)(360,100)
\DashCArc(300,100)(30,0,180){5}
\Text(230,100)[]{\large $b$}
\Text(300,140)[]{\large $H^{\pm}$}
\Text(300,88)[]{\large $u,c,t$}
\Text(370,100)[]{\large $s$}
\end{picture}
\end{center}
\caption{The one-loop diagrams which give contributions to 
$C_7$ and $C_8$ at the electroweak energy scale.  The photon 
or gluon line should be attached appropriately.}
\label{fig:oneloop}
\end{figure}
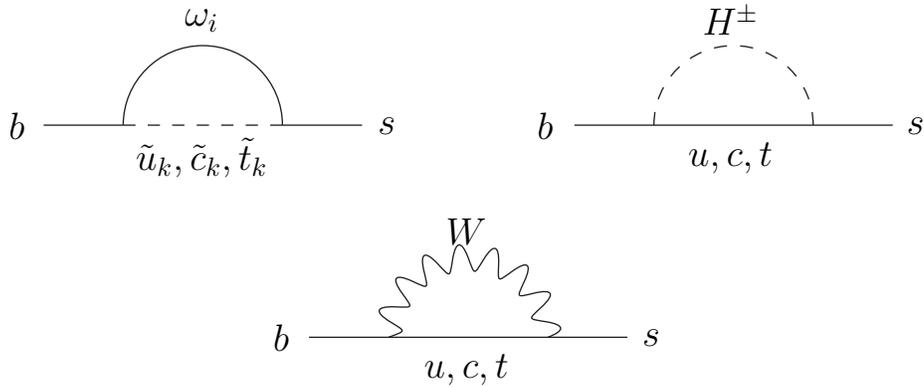

In matching the SSM to the effective Hamiltonian 
at $\mu = M_W$, the coefficients $C_7$ and $C_8$ 
receive contributions from one-loop diagrams which are mediated by 
charginos and charged Higgs bosons with up-type squarks and up-type quarks, 
respectively, as well as by $W$ bosons with up-type quarks.  
The relevant Feynman diagrams are shown in Fig.~\ref{fig:oneloop}.  
The coefficients $C_2(M_W)$, $C_7(M_W)$, and $C_8(M_W)$ are then expressed 
at the leading order (LO) as  
\bsub
\bea
C_2(M_W) &=& 1, 
\\
C_7(M_W) &=& C_7^W(M_W) + C_7^{H^\pm}(M_W) + C_7^\omega(M_W) , 
\\
C_8(M_W) &=& C_8^W(M_W) + C_8^{H^\pm}(M_W) + C_8^\omega(M_W),  
\eea
\esub
where $C_j^W(M_W)$, $C_j^{H^\pm}(M_W)$, and $C_j^\omega(M_W)$ 
$(j=7,8)$ denote, respectively, the contributions from $W$ bosons, 
charged Higgs bosons, and charginos.   From the interaction Lagrangian 
in Eq.~(\ref{chargino lagrangian}), we obtain the chargino 
contributions as 
\bea
C_j^\w(M_W) &=& \sum_{i=1}^2 \frac{M_W^2}{m_{\w i}^2}
	\biggl[ -|C_{R1i}|^2 r_{ui} K_1^j(r_{ui}) 
	- \frac{C_{R1i} C_{L2i}^*}{\r2 \cos\beta} 
	\frac{m_{\w i}}{M_W}r_{ui} K_2^j(r_{ui})              \nonumber \\
	&&	+ \sum_{k=1}^2 \biggl\{ \biggl| C_{R1i} S_{t1k} 
	- \frac{C_{R2i}S_{t2k}}{\r2 \sin\beta} \frac{m_t}{M_W} \biggr|^2
	r_{ki} K_1^j(r_{ki})            \nonumber \\
	&&+ \frac{ C_{L2i}^* S_{t1k}^* }{\r2 \cos\beta} 
	\biggl(C_{R1i} S_{t1k} - \frac{C_{R2i}S_{t2k}}{\r2 \sin\beta}
	\frac{m_t}{M_W}\biggr) \frac{m_{\w i}}{M_W} r_{ki} 
	K_2^j(r_{ki}) \biggr\} \biggr],  \\
 r_{ui} &=& \frac{m_{\w i}^2}{\M_{uL}^2},  \quad 
 r_{ki}=\frac{m_{\w i}^2}{\M_{tk}^2},  \nonumber \\ 
 K_a^7(r) &=& I_a(r)+\frac{2}{3}J_a(r),   \quad 
 K_a^8(r) = J_a(r),  \quad (a=1,2)  \nonumber  
\eea
where the functions $I_a(r)$ and $J_a(r)$ are defined by 
\bsub
\bea
I_1(r) &=& \frac{1}{12(1-r)^4}(2+3r-6r^2+r^3+6r\ln r), \\
I_2(r) &=& \frac{1}{2(1-r)^3}(-3+4r-r^2-2\ln r), \\
J_1(r) &=& \frac{1}{12(1-r)^4}(1-6r+3r^2+2r^3-6r^2\ln r), \\
J_2(r) &=& \frac{1}{2(1-r)^3}(1-r^2+2r\ln r).  
\eea
\esub
The contributions of charged Higgs bosons are given from 
Eq.~(\ref{higgs lagrangian}) by 
\bea
C_j^{\HC}(M_W) &=& -\half r_H \biggl\{ \cot^2\beta 
         \ov K_1^j(r_H) + \ov K_2^j(r_H)\biggr\},  \\
  r_H &=& \frac{m_t^2}{M_{\HC}^2},  \nonumber \\ 
  \ov K_a^7(r) &=& \frac{2}{3}I_a(r)+J_a(r), \quad  
  \ov K_a^8(r) = I_a(r),  \nonumber 
\eea
and the standard $W$-boson contributions lead to 
\bea
C_j^W(M_W) &=& -\frac{3}{2} r_W \ov K_1^j(r_W),  \\
  r_W &=& \frac{m_t^2}{M_W^2}.   \nonumber 
\eea
The coefficients $C_j^W(M_W)$ and $C_j^{\HC}(M_W)$ have real values, 
whereas $C_j^\w(M_W)$ are complex, leading to CP violation.     

An important point is that the chargino contribution can work 
both constructively and destructively.  
The charged Higgs-boson contribution has the same sign as the 
$W$-boson contribution, so that their sum is larger in magnitude 
than the latter alone, $|C_j^W(M_W)+C_j^{\HC}(M_W)|>|C_j^W(M_W)|$.   
On the other hand, owing to a complex value of $C_j^\w(M_W)$, 
the total sum $C_j(M_W)$ may be larger or may be smaller in magnitude 
than $C_j^W(M_W)$ depending on the parameter values.  
It is possible that the conditions $|C_j^\w(M_W)|\sim |C_j^W(M_W)|$ 
and $|C_j(M_W)|\approx |C_j^W(M_W)|$ are satisfied simultaneously.   
In this case CP invariance is violated maximally.  
Even if the branching ratio of $\Bsgamma$ is comparable with 
the SM value, its CP asymmetry may have a large magnitude.  

The Wilson coefficients at $\mu = m_b$ are obtained by solving 
the renormalization group equations  
\bea
\mu \frac{d}{d\mu} C_i(\mu) &=& C_j(\mu) \gamma_{ji}, 
\eea
where $\gamma$ is the anomalous dimension matrix. 
From the LO anomalous dimension matrix the coefficients become 
\bsub
\bea
C_2(m_b) &=& \frac{1}{2} (\eta^{-\frac{12}{23}} + \eta^{\frac{6}{23}} ), 
\\
C_7(m_b) &=& \eta^{\frac{16}{23}} C_7(M_W) + 
 \frac{8}{3} (\eta^{\frac{14}{23}} 
	- \eta^{\frac{16}{23}}) C_8(M_W) + \sum_{i=1}^8 h_i \eta^{a_i}, 
 \\
C_8(m_b) &=& \eta^{\frac{14}{23}} C_8(M_W) + \sum_{i=1}^8 \bar{h}_i \eta^{a_i}, 
\eea
\esub
with $\eta = \alpha_s(M_W)/\alpha_s(m_b)$ which is set for $\eta = 0.56$ 
in the following numerical study.
The constants $h_i, \bar{h}_i$, and $a_i$ are listed  
in Table 1~\cite{buras_misiak_munz_pokorski}. 

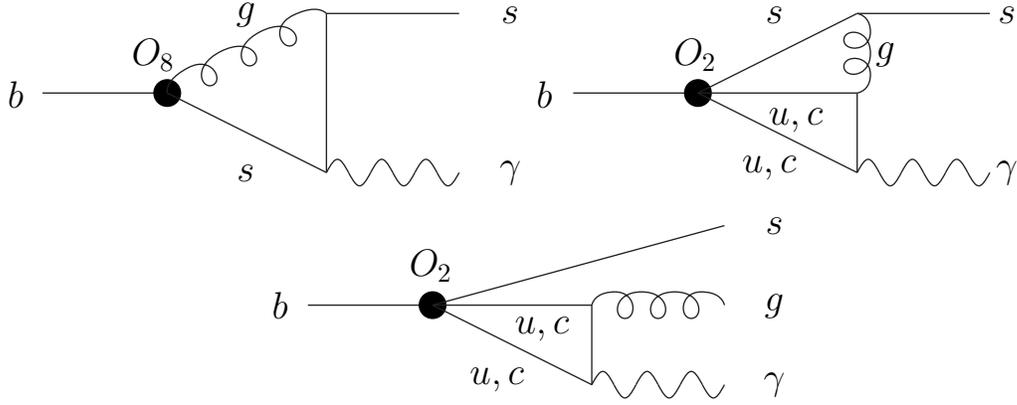
\begin{figure}
\begin{center}
\begin{picture}(400,150)(0,-30)
\Line(23,100)(70,100)
\GOval(70,100)(5,5)(0){0}
\Text(65,115)[]{\large $O_8$}
\Gluon (70,100)(130,130){5}{3}
\Line(70,100)(130,70)
\Line(130,130)(130,70)
\Line(130,130)(180,130)
\Photon(130,70)(180,70){5}{3}
\Text(13,100)[]{\large $b$}
\Text(100,130)[]{\large $g$}
\Text(100,70)[]{\large $s$}
\Text(200,130)[]{\large $s$}
\Text(200,70)[]{\large $\gamma$}
\Line(223,100)(270,100)
\GOval(270,100)(5,5)(0){0}
\Line(270,100)(330,100)
\Line(270,100)(330,130)
\Line (270,100)(330,70)
\Gluon(330,130)(330,100){5}{2}
\Text(270,115)[]{\large $O_2$}
\Line(330,100)(330,70)
\Line(330,130)(380,130)
\Photon(330,70)(380,70){5}{3}
\Text(342,115)[]{\large $g$}
\Text(213,100)[]{\large $b$}
\Text(300,130)[]{\large $s$}
\Text(298,72)[]{\large $u,c$}
\Text(308,90)[]{\large $u,c$}
\Text(388,130)[]{\large $s$}
\Text(388,70)[]{\large $\gamma$}
\Line(123,20)(170,20)
\GOval(170,20)(5,5)(0){0}
\Text(170,35)[]{\large $O_2$}
\Line(170,20)(230,20)
\Gluon(230,20)(280,20){5}{3}
\Line (170,20)(280,50)
\Line(170,20)(230,-10)
\Line(230,20)(230,-10)
\Photon(230,-10)(280,-10){5}{3}
\Text(113,20)[]{\large $b$}
\Text(212,11)[]{\large $u,c$}
\Text(195,-8)[]{\large $u,c$}
\Text(300,20)[]{\large $g$}
\Text(300,50)[]{\large $s$}
\Text(300,-10)[]{\large $\gamma$}
\end{picture}
\end{center}
\caption{The loop diagrams with absorptive parts for $\bsgamma$ 
and $\bsgamma g$, where the blobs stand for the operators 
in Eq. (3.2).  
There are also similar diagrams with the photon line attached 
differently.}
\label{fig:bsg}
\end{figure}

The rate asymmetry in the decay $\Bsgamma$ is attributed 
to decay rate asymmetries for $\bsgamma$ and $\bsgamma g$, 
which are induced by the interferences between the tree diagrams  
and the loop diagrams with absorptive parts as shown in Fig. \ref{fig:bsg}.   
Combining these two contributions, 
the asymmetry in Eq. (\ref{eq:asymmetry}) is given 
by~\cite{kagan_neubert,greub_hurth_wyler}
\bea
\acp &=& \frac{4 \alpha_s(m_b)}{9 |C_7(m_b)|^2} 
	\biggl( \biggl[ \frac{10}{9} - 2z \{ v(z) + b(z, \delta)\} 
		\biggr] {\rm Im} \bigg[	C_2(m_b) C_7^*(m_b) \biggr]
\nonumber \\
	&&
	+ {\rm Im} \biggl[ C_7(m_b) C_8^*(m_b) \biggr]
	+ \frac{2}{3} z b(z, \delta) {\rm Im} 
	\biggl[ C_2(m_b) C_8^*(m_b) \biggr]\biggr), 
\label{eq:cp_asymmetry} \\
v(z) &=& \biggl( 5 + \ln z + \ln^2 z - \frac{\pi^2}{3}\biggr)
	+ \biggl( \ln^2 z - \frac{\pi^2}{3} \biggr) z 
 \nonumber \\
\noalign{\vskip 0.2cm}
	&&~~~~~
	+ \biggl( \frac{28}{9} - \frac{4}{3} \ln z \biggr) z^2 
	+ O(z^3),         
\nonumber \\
\noalign{\vskip 0.2cm}
b(z,\delta) &=& g(z,1) - g(z,1-\delta), 
\nonumber \\
\noalign{\vskip 0.2cm}
g(z,y) &=& \theta(y-4z)
	\biggl\{ (y^2 - 4yz + 6z^2)\ln \biggl( 
	\sqrt{\frac{y}{4z}} + \sqrt{\frac{y}{4z}-1}
				       \biggr)
\nonumber \\ 
\noalign{\vskip 0.2cm}
	&&~~~~~
	- \frac{3y(y-2z)}{4}\sqrt{1 - \frac{4z}{y}}
	\biggr\},   
\nonumber 
\eea
with $z=m_c^2/m_b^2$.   
We have neglected the effect of the standard CP-violating phase 
in the CKM matrix,   
which is known to be small \cite{soares}.  
Owing to the contribution of the three-body decay $b \to s \gamma g$, 
the emitted photon in the inclusive decay process $\Bsgamma$ has 
a continuous energy spectrum. 
The parameter $\delta$ expresses the cut for the photon energy,   
$E_\gamma > (1-\delta)m_b/2$, which is assumed in calculating the asymmetry.  

The theoretical prediction of the branching ratio has large perturbative 
uncertainties at the LO, which are significantly reduced by taking 
into account NLO corrections.  
To the present, these corrections have been calculated for 
the matrix elements at $\mu=m_b$ 
\cite{greub_hurth_wyler,ali_greub} and the anomalous 
dimensions \cite{NLO_anomalous_dimension}.  
For the matching conditions of the Wilson coefficients at 
$\mu = M_W$, calculations have been performed completely 
in the SM~\cite{adel_yao}.     
However, in the SSM, the conditions have only been 
obtained for limited cases \cite{NLO_matching_SSM,2HDM},  
which are not generally applicable.   
We, therefore, calculate the branching ratio of $\Bsgamma$ by using 
the matrix elements and the anomalous dimensions with NLO corrections, 
while for the matching conditions NLO corrections are taken into account 
only for the $W$-boson contributions.  
Our calculations follow formulae given in Ref.~\cite{kagan_neubert2},  
which include also QED corrections.  

%
%
\section{Numerical study}
\clean

We present numerical results of the decay rate asymmetry for
$B \to X_s \gamma$ together with its branching ratio.
Taking into account experimental constraints on the EDMs of the neutron
and the electron, 
we assume that CP violation is induced solely by the phase $\alpha$ 
in Eq. (\ref{eq:cp_sources}) 
and the SU(2) gaugino mass $\wt m_2$ is sufficiently large.
In the following analyses, we fix its value as $\wt m_2=500$ GeV, 
though the asymmetry and the branching ratio 
do not depend much on $\wt m_2$.
The mass ranges $\wt M_{t1}<80$ GeV \cite{lep_stop} and 
$m_{\omega 1}<$ 70 GeV \cite{lep_chargino} 
are experimentally excluded for the lighter $t$ squark $\tilde t_1$ 
and the lighter chargino $\omega_1$, 
respectively, by direct searches for supersymmetry particles. 
In addition, it is imposed that $\tilde t_1$ should be 
heavier than the lightest neutralino from cosmological consideration.
For simplicity, we assume that the soft supersymmetry-breaking masses 
$\wt m_Q$ and $\wt m_U$ have the same value, and define a ratio
$R = \wt m_Q /|A|m_{3/2} =\wt m_U/|A|m_{3/2}$.
For definiteness, measurable parameters are fixed at the $b$-quark mass scale 
as the followings:
$m_b=4.8$ GeV, $z= m_c^2/m_b^2=(0.29)^2$, 
$|V_{ts}^{\ast}V_{tb}/V_{cb}|=0.97.$
The energy cutoff parameter $\delta$ for the photon 
is taken to be 0.99, though the decay rate 
asymmetry is not changed so much by the choice of its value.

\begin{figure}[t]
	\begin{center}
	\leavevmode\psfig{figure=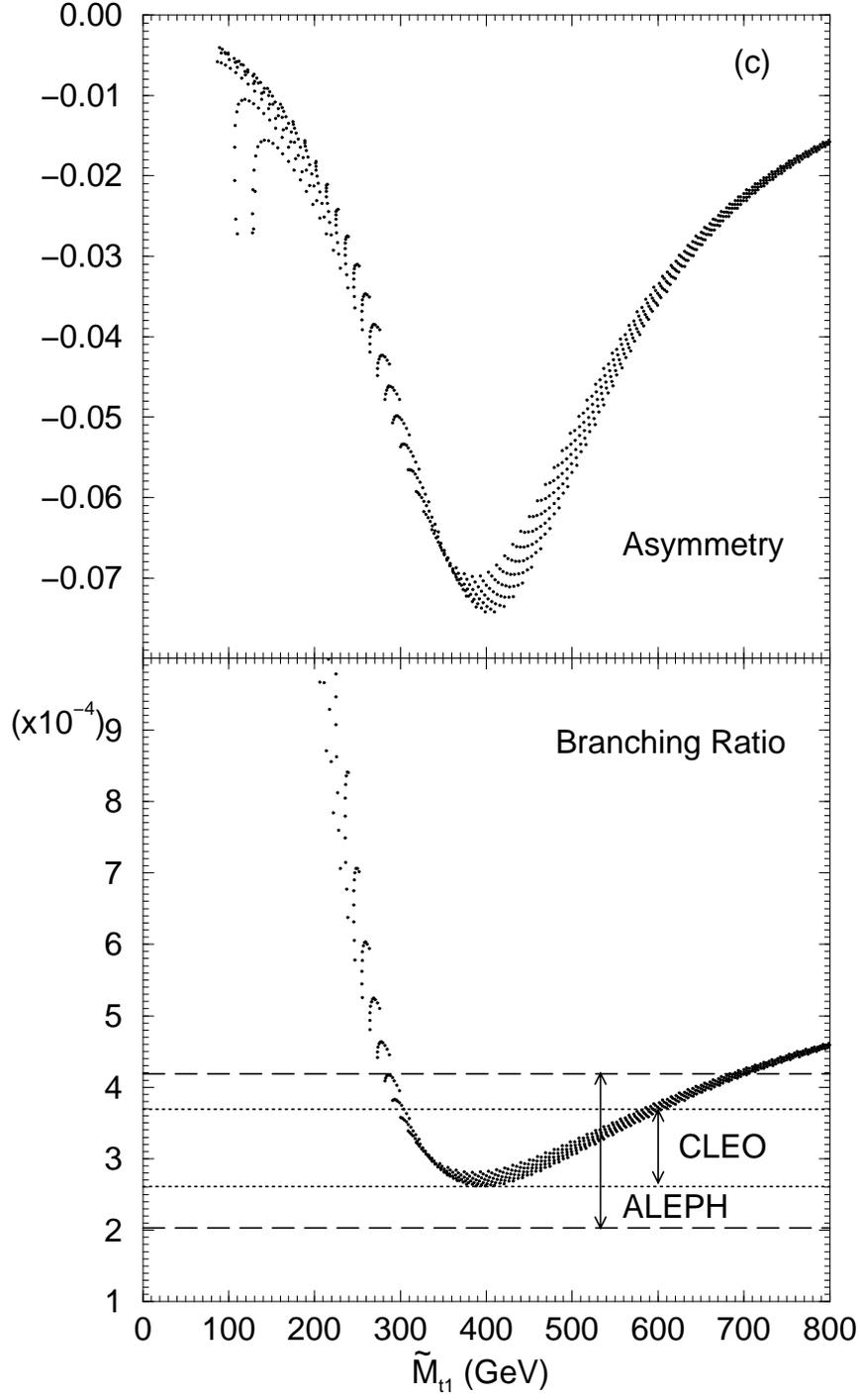,height=19cm}
	\end{center}
	\caption{The decay rate asymmetry (upper) and the branching ratio
(lower) of $B \to X_s \gamma$ for $R=1$, $\alpha=\pi/4$, $m_H=100$ GeV, 
and $M_{H^{\pm}}=200$ GeV.  
(a) $\tan \beta=2$, (b) $\tan \beta=10$, (c) $\tan \beta=30$.}
	\label{fig:asy_tanb}
	\end{figure}

The decay rate asymmetry and the branching ratio 
are shown in Fig. \ref{fig:asy_tanb} as functions of the lighter 
$t$-squark mass $\wt M_{t_1}$ for (a) $\tan\beta =2$, (b) $\tan \beta=10$,
and (c) $\tan \beta =30$.
We set the CP-violating phases for $\alpha=\pi/4$ and $\theta=0$.
The mass parameters for the charged Higgs boson and the Higgsino
are taken for $M_{H^\pm}= 200$ GeV and $|m_H|=100$ GeV.
The ratio  $R$ is set for $R = 1$.
The parameters $c$ and $|A|m_{3/2}$ are scanned over $0.1 \leq c \leq 1$ 
and $|A|m_{3/2} \leq$ 1 TeV.
The experimental upper and lower bounds of the branching 
ratio~\cite{cleo,aleph} are also indicated.  
For $\tan\beta=10$, the branching ratio lies within the experimental bounds 
by ALEPH in the range 100 GeV$\lsim \mstone \lsim$400 GeV, where  
the asymmetry has a value $0.02 \lsim |\acp| \lsim 0.07$.
The peaks of the asymmetry and the branching ratio are both roughly
given at $\mstone \simeq 200$ GeV.
For $\tan\beta=2$ and $\tan\beta=30$, the branching ratio is consistent with 
its experimental bounds in the ranges 100 GeV$\lsim \mstone \lsim$200 GeV 
and 300 GeV$\lsim \mstone \lsim$700 GeV, 
where the asymmetry has values $0.01 \lsim |\acp| \lsim 0.06$ and 
$0.02 \lsim |\acp| \lsim 0.075$, respectively.
The peaks of the asymmetry and the branching ratio are  
roughly at the same value of $\mstone$, which increases with $\tan \beta $.
The maximal value of the asymmetry does not depend
significantly on $\tan \beta$. 
As $\tan\beta$ increases, larger masses for $\tilde t_1$ become 
allowed by the experimental constraints.

\begin{figure}[t]
	\begin{center}
	\leavevmode\psfig{figure=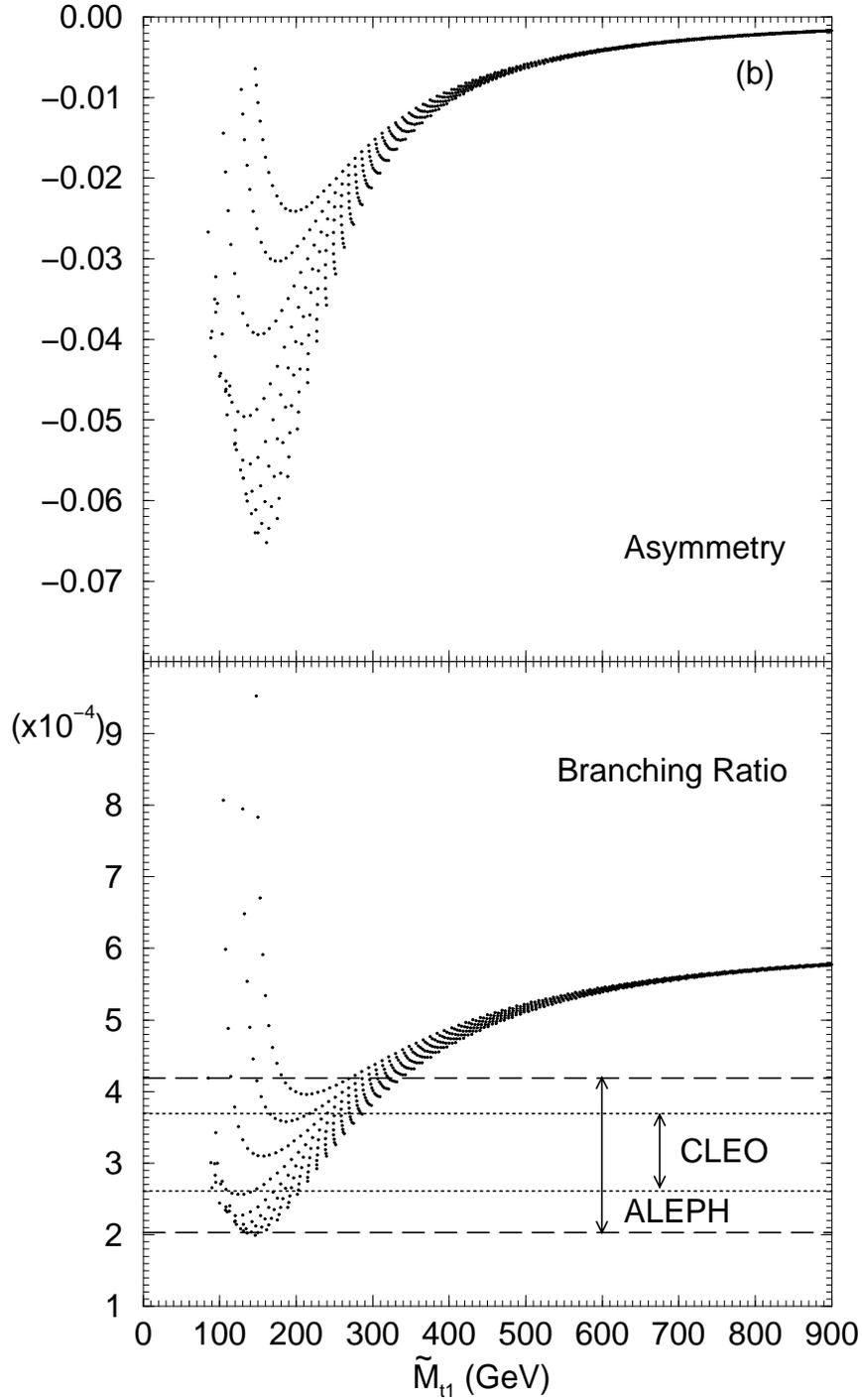,height=19cm}
	\end{center}
	\caption{The decay rate asymmetry (upper) and the branching ratio
(lower) of $B \to X_s \gamma$ for $\tan\beta=10$, $\alpha=\pi/4$, 
$m_H=100$ GeV, and $M_{H^{\pm}}=200$ GeV.  
(a) $R=0.5$, (b) $R=2$.}
	\label{fig:asy_r}
	\end{figure}

We show in Fig. \ref{fig:asy_r} the asymmetry and the branching ratio 
for (a) $R=0.5$ and (b) $R=2$, with $\tan \beta =10$.
The other parameters are taken for the same values as 
in Fig. \ref{fig:asy_tanb}.
The allowed region for the mass of $\tilde t_1$ is given by 
150 GeV$\lsim \mstone \lsim$500 GeV for $R=0.5$, where 
the asymmetry is in the range $0.02 \lsim |\acp| \lsim 0.075$.
For $R=2$, under the condition that $\mstone$ remain the same,
the amount of mixing for $\tilde t_L$ and $\tilde t_R$ varies 
very much with $c$, 
leading to strong $c$-dependencies of the asymmetry and the branching 
around their peaks.  
The asymmetry becomes $0.01 \lsim |\acp| \lsim 0.065$.

\begin{figure}[t]
	\begin{center}
	\leavevmode\psfig{figure=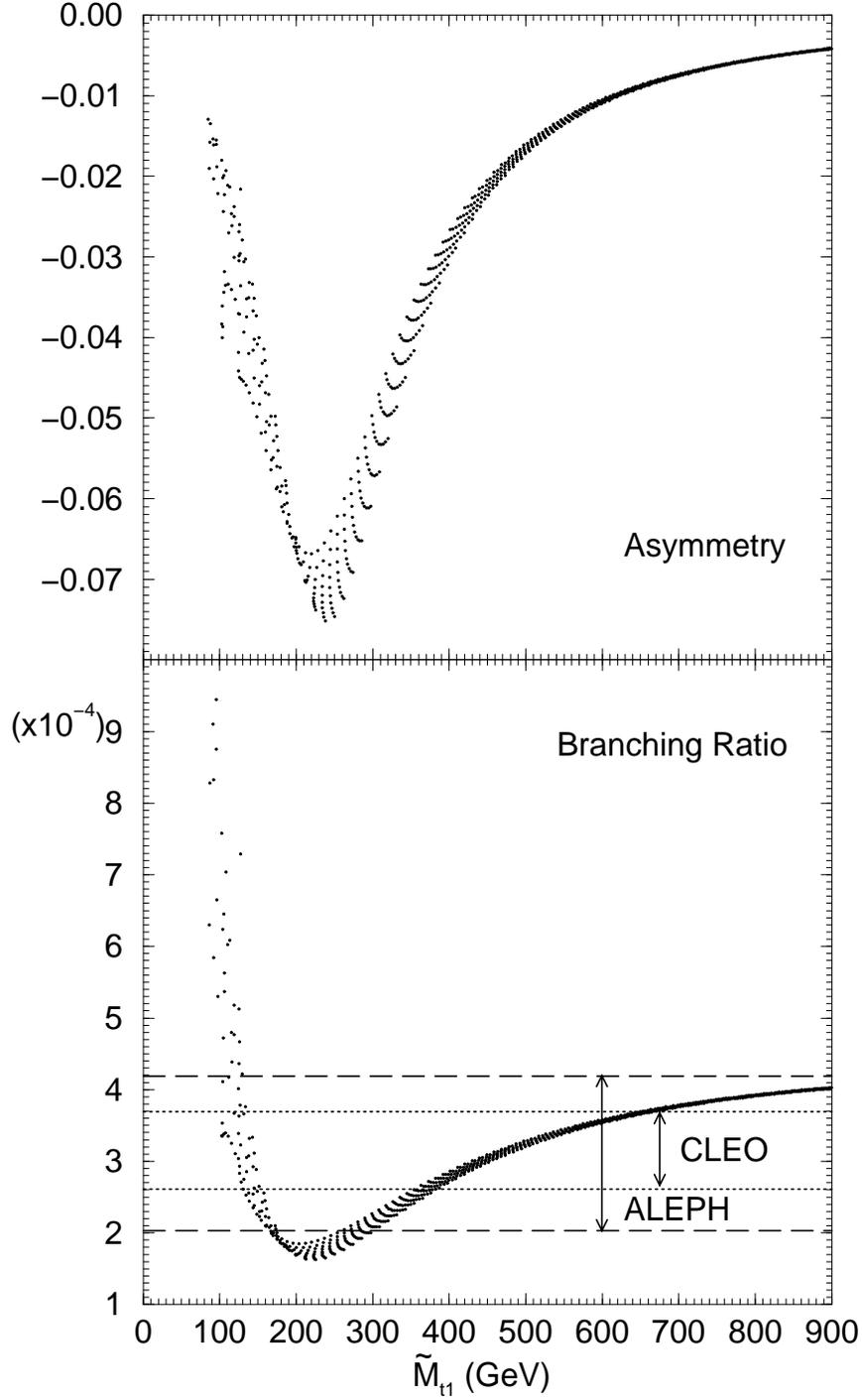,height=19cm}
	\end{center}
	\caption{The decay rate asymmetry (upper) and the branching ratio
(lower) of $B \to X_s \gamma$ for $M_{H^\pm}= 500$ GeV, $R=1$, 
$\tan\beta=10$, $\alpha=\pi/4$, and $m_H=100$ GeV.}
	\label{fig:asy_higgs}
	\end{figure}

In Fig. \ref{fig:asy_higgs}, the asymmetry and the branching ratio are
shown for $M_{H^{\pm}}$= 500 GeV, with $\tan \beta =10$.
We take the other parameters for the same values as 
in Fig. \ref{fig:asy_tanb}.
Since the charged Higgs-boson contribution to $b \to s \gamma$
decreases as $M_{H^{\pm}}$ becomes larger, the branching ratio
is smaller than that for $M_{H^{\pm}}$= 200 GeV.
As a result, a large value of $\mstone$ becomes allowed 
by the experimental constraints for the branching ratio,
whereas for 150 GeV$\lsim\mstone\lsim$300 GeV the branching ratio 
is too small.  
The asymmetry is given by $|\acp|\lsim 0.06$ in the experimentally 
allowed range, and does not vary much with $M_{H^{\pm}}$.  

\begin{figure}[t]
	\begin{center}
	\leavevmode\psfig{figure=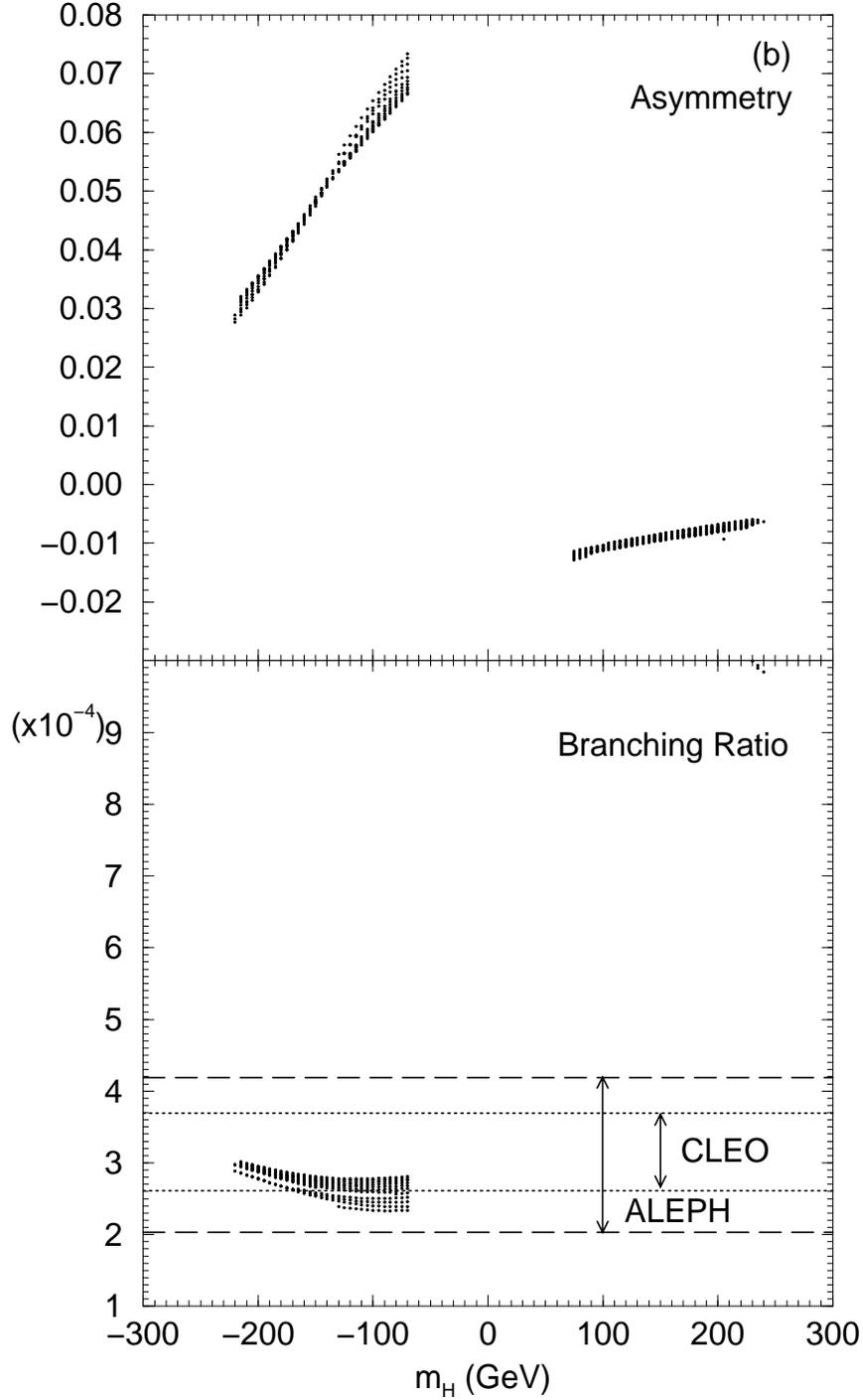,height=19cm}
	\end{center}
	\caption{The decay rate asymmetry (upper) and the branching ratio
(lower) of $B \to X_s \gamma$ as functions of $m_H$ for 
 $R=1$, $\tan\beta=10$, and $M_{H^\pm}= 200$ GeV.
The mass of $\tilde t_1$ is taken for 195 GeV $\leq \mstone \leq$ 205 GeV.  
(a) $\alpha=\pi/4$, (b) $\alpha=3\pi/4$. }
	\label{fig:asy_phase}
	\end{figure}

We plot in Fig. \ref{fig:asy_phase} the asymmetry and the branching ratio 
as functions of the Higgsino mass parameter $m_H$ for 
(a) $\alpha=\pi/4$ and (b) $\alpha=3\pi/4$, with $\tan\beta=10$.
The mass of $\tilde t_1$ is kept for 195 GeV$\leq \mstone \leq$205 GeV.
The phase $\theta$ is taken for 0 or $\pi$, corresponding to 
positive or negative values for $m_H$, respectively.
The other parameters are taken for the same values as 
in Fig. \ref{fig:asy_tanb}.
The ranges for smaller values of $|m_H|$, where no plot is given, are excluded 
by the experiments since $m_{\omega 1}$ becomes less than 70 GeV.
In the ranges for larger values of $|m_H|$ with no plot, 
the lightest neutralino is heavier than $\tilde t_1$.   
For $\alpha=\pi/4$, in the range with $m_H>0$ the chargino contributions  
to $C_7$ and $C_8$ are added destructively to the standard $W$-boson 
and the charged Higgs-boson contributions.    
On the other hand, in the range with $m_H<0$ the formers are  
added constructively to the latters, making the branching ratio too large.   
Such a relative relation depends on not only $\theta$ but also $\alpha$.
For $\alpha=3\pi/4$, the branching ratio is too large for $m_H > 0$ 
while within the experimental bounds for $m_H <0$.
For either value of $\alpha$, the asymmetry is given by 
$0.02\lsim|\acp|\lsim 0.075$ in the experimentally allowed ranges.

Summarizing these numerical results, the absolute value of the decay rate 
asymmetry in the SSM is larger than 0.02 in wide parameter regions 
where the branching ratio is consistent with its experimental bounds. 
If $M_{H^{\pm}}$ is of order 100 GeV, the charged Higgs-boson 
contribution to $\Bsgamma$ is not small \cite{2HDM}.
Since the experimental values for the branching ratio are almost 
consistent with the SM prediction, the chargino-squark loop
diagram has to give a large destructive contribution in order to cancel out 
the large charged Higgs-boson contribution.
Therefore, as long as the CP-violating phase $\alpha$ is not suppressed,
the chargino contribution induces a sizable CP asymmetry.
On the other hand, if the charged Higgs boson is heavy, 
its contribution becomes negligible.  
Still, as shown in Fig. \ref{fig:asy_higgs},
there are parameter regions where the asymmetry can be 
sizable without conflicting with the measured branching ratio.   
Because it is possible that the chargino contributions make 
the Wilson coefficients $C_7$ and $C_8$ complex while 
their absolute values being kept unchanged.   

%
%
%
\section{Discussions}
\clean

The decay rate asymmetry for $\Bsgamma$ 
could be measured at B factories.   It is expected there that 
$B\ov B$ pairs of order $10^{8}$ will be  produced per year, 
among which approximately half are pairs of $B^+B^-$.  
Since $B^0$ and $\ov B^0$ may make the transition to each other 
before their decays, $B^+B^-$ pairs are suitable for the measurement 
of the asymmetry discussed in this paper.  
For estimating measurability 
we tentatively assume that the tagging efficiency 
of $\Bsgamma$ by detecting the photon is around 0.3.  
A distinction between $B^+$ and $B^-$ can be made by examining 
the strange particle accompanied with the photon  
or the semileptonic decay of the opposite $B$ meson.  
We assume these tagging efficiencies to be around 0.1 each.  
Then, a detectable number of $\Bsgamma$ with the charge of $B$ being specified 
is roughly of order $10^3$.  The asymmetry is measurable 
to a level of a few percent.  The decay rate asymmetry induced by the SSM 
may be well within reach of B factories.    

\begin{figure}[t]
	\begin{center}
	\leavevmode\psfig{figure=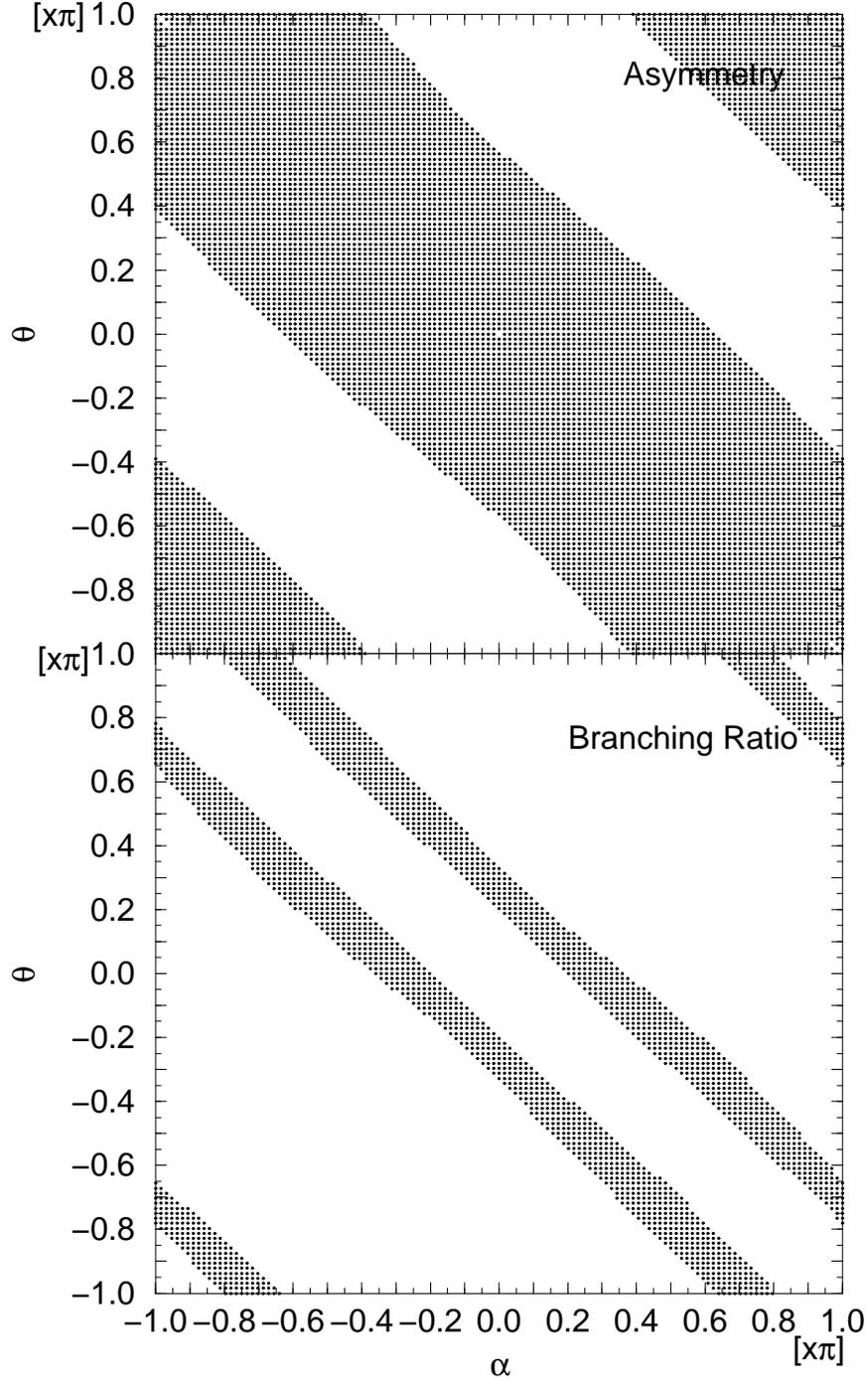,height=19cm}
	\end{center}
	\caption{The regions (dotted) in the $(\alpha,\theta)$ plane for 
     $|\acp|\geq 0.02$ (upper) and for the branching ratio allowed 
     by the ALEPH experiment (lower).  The parameter values are set 
    for $\mgaugino_2=200$ GeV, $|m_H|=100$ GeV, 
195 GeV $\leq \mstone \leq$ 205 GeV, 
 $R=1$, $\tan\beta=10$, and $M_{H^\pm}= 200$ GeV.}
	\label{fig:asy_extra}
	\end{figure}

The SSM has several new contributions to the EDMs 
of the neutron and the electron, yielding nontrivial     
constraints on the model parameters.  
In general, it is necessary that CP-violating phases 
are suppressed or supersymmetry particles are heavy.  
Nevertheless, as we have shown, there are sizable 
regions of parameter space in which the decay rate asymmetry 
of $\Bsgamma$ becomes large.  
On the other hand, it may be possible \cite{ibrahim_nath} that 
the EDMs are suppressed by cancellation among different 
contributions while each contribution is not small.  
In this case, unsuppressed CP-violating 
phases with light supersymmetry particles could be allowed.  
In Fig. \ref{fig:asy_extra}, discarding 
the constraints from the EDMs, we show parameter ranges for the 
CP-violating phases $\alpha$ and $\theta$ 
in which the asymmetry is larger than 0.02 and those in which 
the branching ratio lies within the experimental bounds.   
The conditions are satisfied in dotted regions 
except the vicinities of CP-invariant points, i.e. $\sin\theta=\sin\alpha=0$, 
for the asymmetry.  
The SU(2) gaugino mass is taken for $\mgaugino_2=200$ GeV and   
the $\tilde t_1$ mass is kept for 195 GeV$\leq\mstone\leq$205 GeV, 
with $\tan\beta=10$.  The other parameters are set for the same values  
as in Fig. \ref{fig:asy_tanb}.  
We can see that the asymmetry is generally 
large for the parameter values consistent with the branching ratio,  
similarly to the results of our previous analyses.   
Maximal values of the asymmetry are of order 0.1.    
If the small values of the EDMs are due to the cancellation, 
the decay rate asymmetry is also likely to be large. 
It should be noted that 
the CP-invariant points are not allowed by the branching ratio 
in Fig. \ref{fig:asy_extra}.  
For discussing the radiative $B$-meson decay in the SSM, 
neglect of CP-violating phases could lead to incorrect conclusions.  

The decay rate asymmetry for $\Bsgamma$ in the SSM  
has also been studied recently under different assumptions.  
In Refs. \cite{kagan_neubert,chua_he_how_kim_ko_lee},   
assuming flexibility of mass-squared matrices for squarks,     
the contributions by gluino-mediated and chargino-mediated 
diagrams are discussed and large values for the asymmetry are obtained.   
In the ordinary scheme based on supergravity and GUTs, 
however, the gluino contribution becomes much smaller 
than the chargino contribution, owing to a 
small mixing between $\tilde b_L$ and $\tilde b_R$.  
In Ref. \cite{goto_keum_nihei_okada_shimizu}, based on supergravity and GUTs, 
the chargino contribution is studied in a minimal model.  
Assuming universal values for soft-supersymmetry-braking   
parameters at the GUT scale, small values for the asymmetry  
are obtained, which is traced back to small CP-violating phases 
at the electroweak scale to satisfy the constraints from 
the EDMs \cite{matsumoto_arafune_tanaka_shiraishi}.  
However, if the universality or the minimality at the GUT scale is lifted, 
there is more freedom for the SSM parameters and  
their values adopted in our analyses could be well allowed.  

We have studied the radiative $B$-meson decay, concentrating 
its decay rate asymmetry induced by a new source of CP violation 
in the SSM.  
The already available experiments give both directly and 
indirectly nontrivial constraints on the SSM parameters.  
In sizable regions of the parameter space allowed by 
these experiments, the asymmetry 
is predicted to have a magnitude larger than 0.01.  
Such a magnitude of the asymmetry is larger than the SM prediction.  
Moreover, it may be possible to detect the asymmetry at B factories.  
  
%
%
%
%
\section*{Acknowledgments}

The authors thank M. Yamauchi for discussions on experiments at 
B factories.  
This work is supported in part by Grant-in-Aid for Scientific 
Research from the Ministry of Education, Science and Culture of Japan.

%
%
%
%
%
%

\newpage
\begin{table}
\begin{center}
\begin{tabular}{c c c c c c c c c}
\hline
 $i$   & 1 & 2 & 3 & 4 & 5 & 6 & 7 & 8 \\
\hline
   & & & & & & & & \\
 $a_i$  & $\frac{14}{23}$ & $\frac{16}{23}$ & $\frac{6}{23}$ & $-\frac{12}{23}$ 
         & 0.4086 & $-0.4230$ & $-0.8994$ & 0.1456 \\
   & & & & & & & & \\
 $h_i$  & $\frac{626126}{272277}$ & $-\frac{56281}{51730}$ & $-\frac{3}{7}$ 
       & $-\frac{1}{14}$ & $-0.6494$ & $-0.0380$ & $-0.0186$ & $-0.0057$ \\
   & & & & & & & & \\
 $\bar h_i$  & $\frac{313063}{363036}$ & 0 & 0 & 0 
       & $-0.9135$ & 0.0873 & $-0.0571$ & 0.0209   \\
   & & & & & & & & \\
\hline
\end{tabular}
\end{center}
\caption{The values of $a_i$, $h_i$, and $\bar h_i$ 
in Eq. (3.9).  }
\label{tab1}
\end{table}
%

\newpage

\newpage

\newpage
	\begin{figure}[t]
	\begin{center}
	\leavevmode\psfig{figure=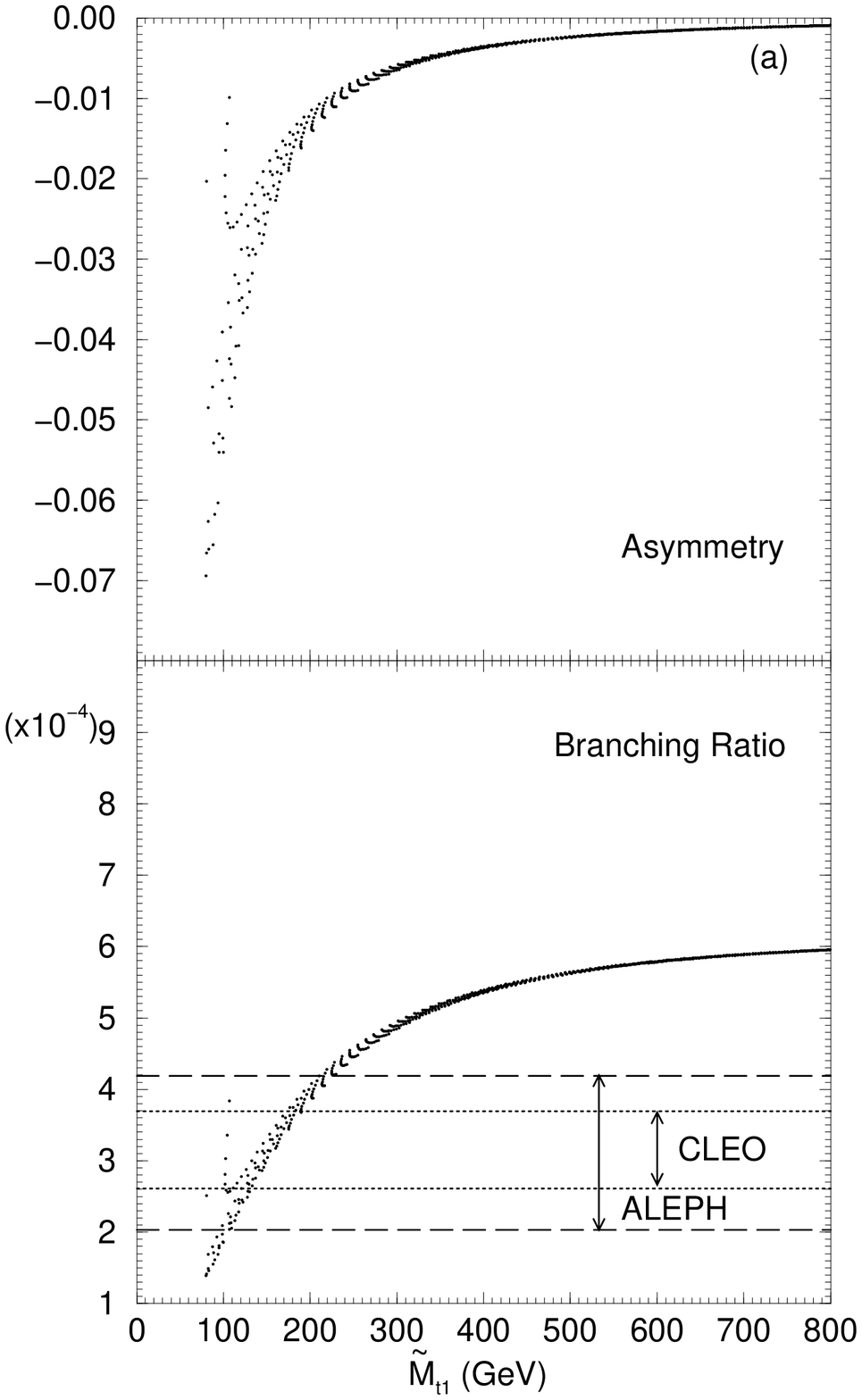,height=19cm}
	\end{center}
	\end{figure}
\newpage
	\begin{figure}[t]
	\begin{center}
	\leavevmode\psfig{figure=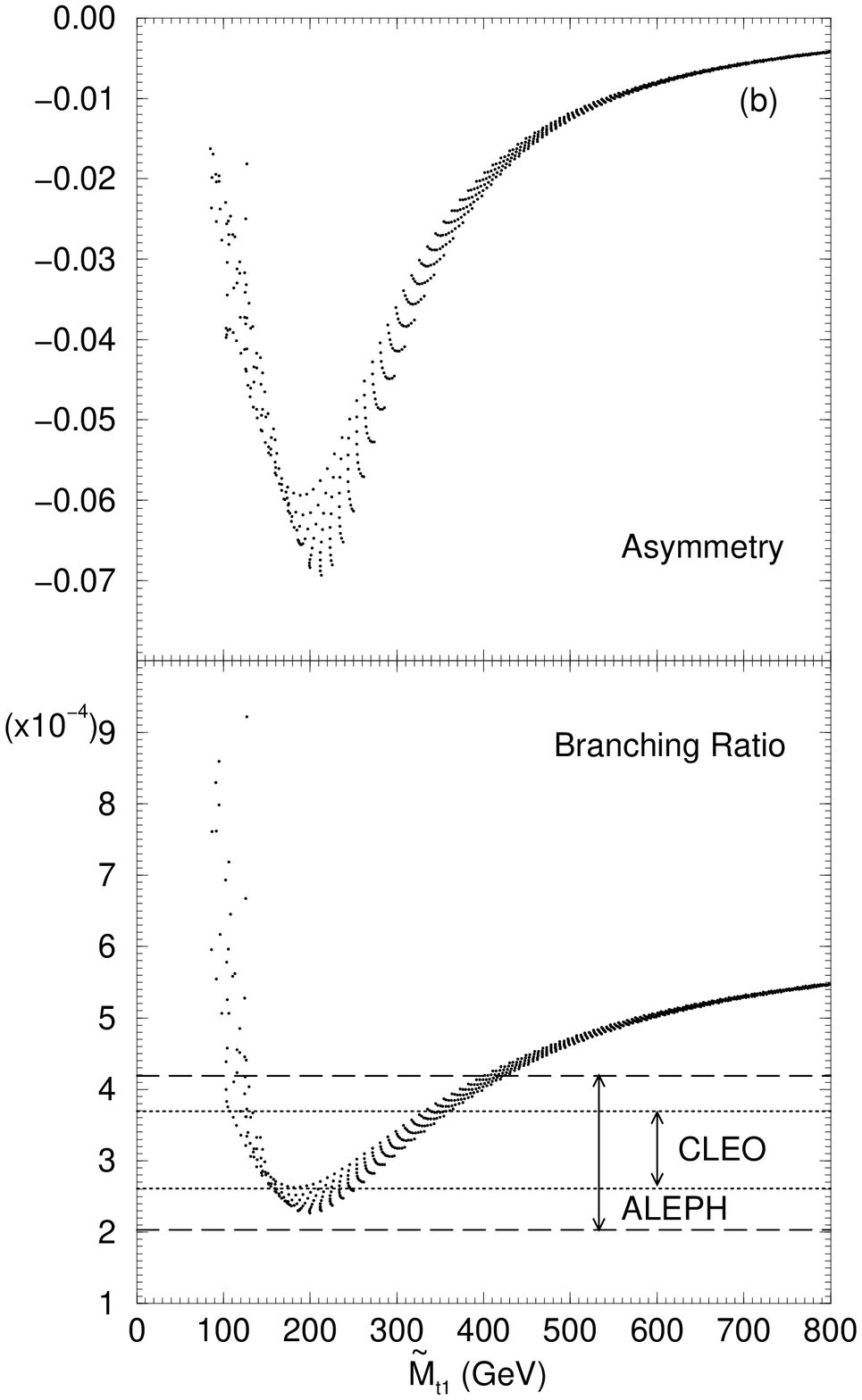,height=19cm}
	\end{center}
	\end{figure}
\newpage
\newpage
	\begin{figure}[t]
	\begin{center}
	\leavevmode\psfig{figure=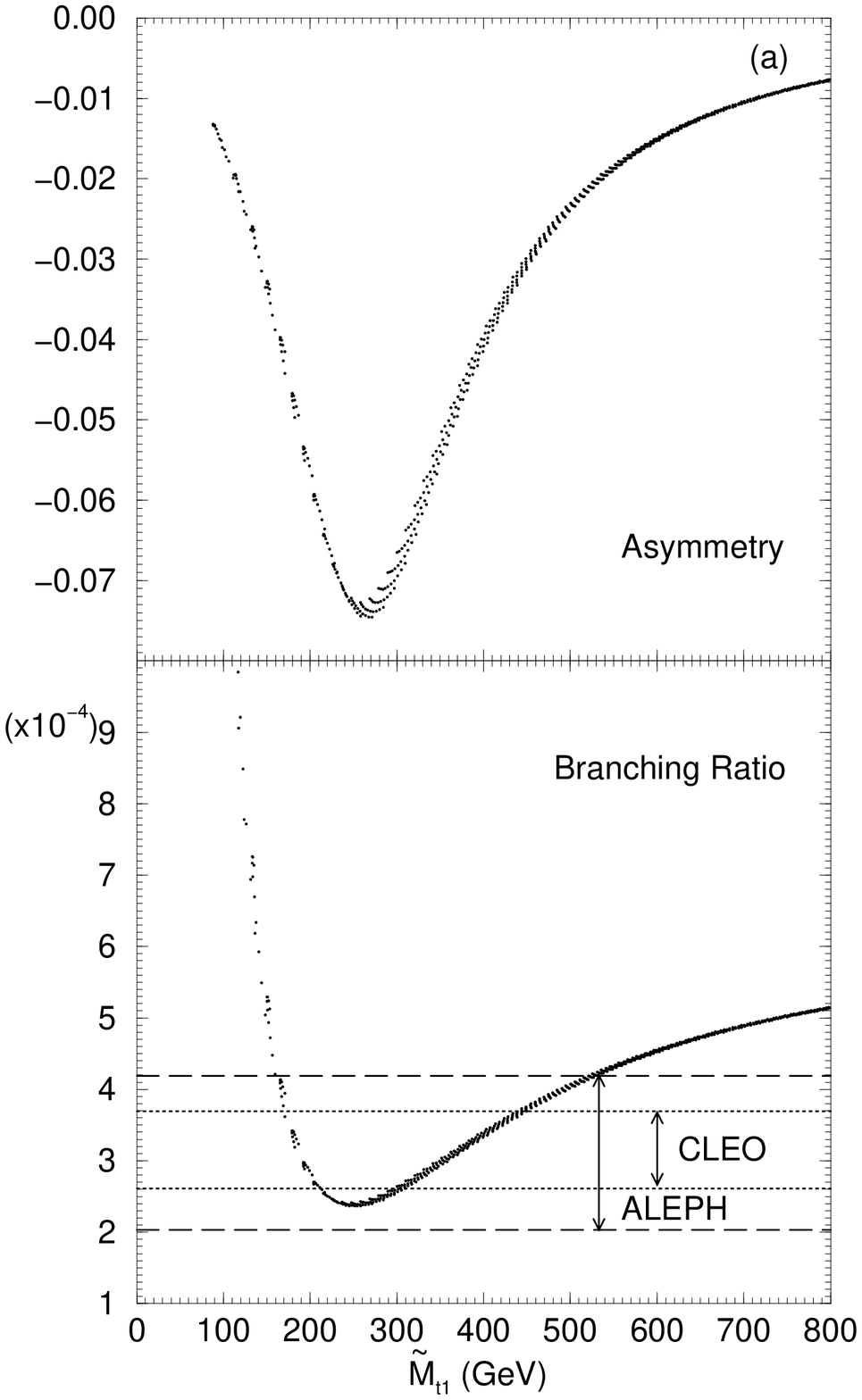,height=19cm}
	\end{center}
	\end{figure}
\newpage
\newpage
\newpage
	\begin{figure}[t]
	\begin{center}
	\leavevmode\psfig{figure=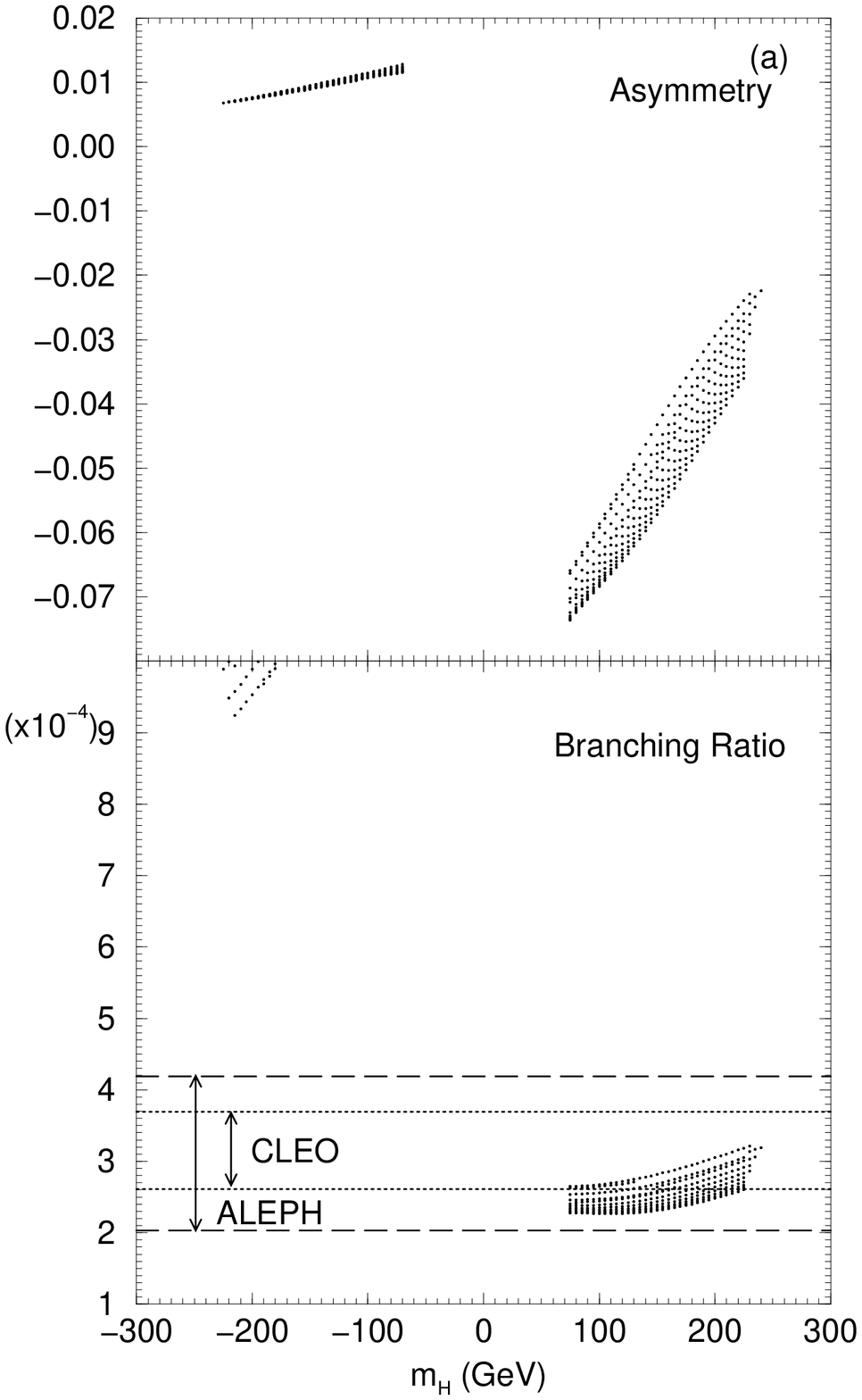,height=19cm}
	\end{center}
	\end{figure}
\newpage
\newpage
\end{document}